\documentstyle[aps,prl,preprint]{revtex}

\begin{document}
\title{Shot noise in linear macroscopic resistors}
\author{$^{1}$G. Gomila, $^{2}$C. Pennetta,$^{2}$L. Reggiani, $^{3}$M. Sampietro, $%
^{3}$G. Ferrari and $^{3}$G. Bertuccio}
\address{$^{1}$Department d'Electronica and Research Centre for Bioelectronics and\\
Nanobioscience, Universitat de Barcelona, Edifici Modular C/ Josep Samitier 1-5, 08028\\
Barcelona, Spain.\\
$^{2}$INFM-National Nanotechnology Laboratory and Dipartimento di\\
Ingegneria dell'Innovazione, Universit\`{a} di Lecce, Via Arnesano s/n,\\
73100 Lecce, Italy.\\
$^{3}$Dipartimento di Elettronica ed Informazione, Politecnico di Milano,\\
Pzza Leonardo da Vinci 32, 20133 Milano, Italy.}
\maketitle

\begin{abstract}
We report on a direct experimental evidence of shot noise in a linear
macroscopic resistor. The origin of the shot noise comes from the
fluctuation of the total number of charge carriers inside the resistor
associated with their diffusive motion under the condition that the
dielectric relaxation time becomes longer than the dynamic transit time.
Present results show that neither potential barriers nor the absence of 
inelastic scattering are necessary to observe 
shot noise in electronic devices.
\end{abstract}

\draft


Nyquist noise and shot noise are the two prototypes of current noise
displayed by electronic devices. Nyquist noise, originally found by Johnson
in resistors\cite{johnson28}, is displayed by all electronic devices at
thermal equilibrium and is associated with the equilibrium thermal
fluctuations.The spectral density of the current fluctuations of Nyquist
noise is white and given by\cite{nyquist28} 
\begin{equation}
S_{I}^{Nyquist}=\frac{4k_{B}T}{R}  \label{Nyquist}
\end{equation}
where $k_{B}$ is the Boltzmann constant, $T$ the temperature and $R$ the
linear resistance. Shot noise, originally found in saturated vacuum tubes,%
\cite{schottky18} is displayed under non-equilibrium conditions and is
associated with the discreteness of the electric charge. Its current
spectral density at low frequency is usually given by \cite{schottky18} 
\begin{equation}
S_{I}^{shot}=2qI  \label{Schottky}
\end{equation}
where $I$ is the average current and $q$ the carrier charge. Shot noise is
routinely found in many solid-state electronic devices like tunnel and
Schottky diodes, p-n junctions\cite{ziel86}, and more recently in mesoscopic
structures.\cite{blanter00} The existing claiming is that shot noise can be
observed in electronic devices provided there exists an internal potential
energy barrier\cite{ziel86} or in the absence of inelastic scattering. \cite
{blanter00,shimizu92,dejong95} Here, we report on a direct experiment
evidence of shot noise in a linear macroscopic resistor, which does not
satisfy the above requeriments.

In a macroscopic linear resistor, the possibility to observe shot noise is
conditioned to the fact that the electrical charge can pile up inside the
device \cite{vliet56,buttiker95,gomila00} or, conversely, to the fact that
the instantaneous number of free carriers inside the sample can fluctuate in
time.\cite{landauer93} This possibility can be accomplished when the
dielectric relaxation time of the material, i.e. the time required for a
charge fluctuation to vanish, $\tau _{d}=\epsilon \epsilon _{0}\rho $,
becomes longer than the dynamic transit time, i.e. the time a particle lasts
to cross the sample at its drift velocity, $\tau _{T}=L^{2}/(\mu V)$. Here $%
\epsilon _{0}$ is the vacuum permittivity, $\epsilon _{r}$ the relative
static dielectric constant of the material, $\rho $ its resistivity, $L$ its
length, $\mu $ its mobility, and $V$ the applied voltage. Under such a
condition the long range Coulomb interaction does not induce correlations
between current fluctuations, thus, charge neutrality of the device can be
violated and shot noise can be observed. In order to fulfil this constraint (%
$\tau _{T}\ll \tau _{d}$), numerical estimates indicate that the choice of
the sample is limited to highly resistive materials like semi-insulating
semiconductors. In relatively good conductors, like metals or highly doped
semiconductors, the above constraint is hard, if not impossible, to be
achieved under realistic experimental conditions.

In the present work we have considered a resistor made of a $2$ mm thick
semi-insulating CdTe semiconductor embedded between two gold plates of $%
2\times 10$ mm$^{2}$. The choice of CdTe has been motivated by the fact that
this semiconductor material allows for a high degree of compensation\cite
{strauss97}, thus presenting the desired semi-insulating property .
Moreover, it displays a linear velocity-field characteristics up to several
kV/cm at room temperature,\cite{strauss97} thus allowing to apply
considerable high voltages without the presence of hot-electron effects.
Furthermore, the use of metal-semiconductor contacts is motivated by the
fact that metals on semi-insulating materials exhibit the required nearly
perfect ohmic behaviour in a wide range of voltages, since the carrier
density at the interface imposed by the contact is of the same order of
magnitude as the free carrier density of the semi-insulating material.\cite
{gomila03} This fact avoids the presence of spurious space charge effects.
Taken, for example, at T =323 K, the device displays an almost symmetric
linear current-voltage ($I-V$) characteristics between 50 and + 50 V. The
forward characteristics is shown in Fig. \ref{Fig1} in a log-log scale
showing the linear behaviour from the lowest to the largest voltage bias. A
best fit to the experiments gives a resistance $R_{323K}=0.233$ G$\Omega $,
which implies a resistivity $\rho =0.233$ G$\Omega $ cm. From these
parameters, and by using\cite{strauss97} $\epsilon _{r}=12$, the dielectric
relaxation time for the material is $\tau _{d}\sim 0.3$ ms. Therefore, by
extracting a mobility corresponding to holes $\mu =50$ cm$^{2}$/(Vs) from
the meaurement of the cut-off frequency in the noise spectra in the shot
noise region \cite{sampietro00} (the sample is known to be p-type\cite
{cavallini03}), and for an applied bias of $10$ V, the transit time is $\tau
_{T}\sim 0.08$ ms , thus making accessible the shot noise condition ($\tau
_{T}\ll \tau _{d}$ ) for applied bias above a few tens of Volts.

Current noise experiments have been performed on the CdTe resistor by means
of the correlation technique implemented on a state of the art noise
spectrum analyzer able to probe noise levels as low as $10^{-30}$ A$^{2}$/Hz
and to reach frequencies up to $10^{5}$ Hz at room temperature\cite
{sampietro99}. The characteristics of the measurement set up allow reaching
the extremely low current noise levels present in semi-insulating materials
and to cover a wide enough range of frequencies to get rid of the $1/f$
contribution that may hide the presence of the shot noise plateau. Figure 
\ref{Fig2} reports the spectral density of the current fluctuations measured
at $T=323$ K for different values of the current. At thermal equilibrium
(i.e. zero current) the spectrum is white and takes a value in agreement
with Nyquist noise, Eq.(\ref{Nyquist}), as should be. At increasing currents
the spectrum becomes current and frequency dependent. It displays a $1/f$
region at low frequencies, followed by a plateau at intermediate
frequencies, and a cut-off region at the highest frequencies.

The values of the plateaux for the different current values extracted by a
best fit function that includes the plateaux and the cut-off regions are
reported as a function of current in Fig. \ref{Fig3}. The dependence of the
current spectral density at the plateaux on current is found to exhibit
three distinct behaviours. At the lowest currents the spectral density keeps
a constant value corresponding to Nyquist noise. At intermediate currents
the spectral density exhibits a sharp increase (cross-over region) until
approaching an asymptotic linear increase with current that merges with the
full shot noise behaviour. We note that the onset of shot noise occurs at a
current around $100$ nA, which corresponds to a voltage $\sim 20$ V, in
agreement with the voltage estimated above. These experiments prove that a
linear macroscopic resistor can display shot noise under the condition that
the dynamic transit time is shorter than the dielectric relaxation time of
the sample. It is also worth remarking that the cross-over between Nyquist
and shot noise is found to depart from the standard coth-like behaviour
observed in many solid-state devices (dotted line in Fig. \ref{Fig3}).

In order to show that the shot noise displayed by the CdTe resistor is not
related to the presence of some sort of potential energy barrier or to the
absence of inelastic scattering, we provide a quantitative theoretical
interpretation of the experiments by means of a unipolar drift-diffusion
noise model recently developed by two of the Authors.\cite{gomila00} The
model assumes that charge transport and current fluctuations are due to the
drift and diffusion of free carriers in a spatially homogeneous (in average)
electric field. The fact that the resistor is macroscopic is implicitly
accounted for by assuming that the free carriers are in local thermal
equilibrium with the lattice at the bath temperature due to the energy
exchange with the lattice phonons. In the model,{\em \ the only source of
noise is that related to the diffusive motion of the carriers} The
analytical solution of the model can be obtained, giving for the low
frequency current spectral density the following relation\cite{gomila00}

\begin{equation}
S_{I}=\frac{4k_{B}T}{R}\left( 1+s_{I}^{ex}\right)   \label{SID}
\end{equation}
where 
\begin{eqnarray}
s_{I}^{D,ex} &=&\frac{\left( \lambda _{2}^{2}-\lambda _{1}^{2}\right) \left(
e^{\lambda _{1}}-1\right) \left( e^{\lambda _{2}}-1\right) }{2\lambda
_{1}^{2}\lambda _{2}^{2}\left( e^{\lambda _{1}}-e^{\lambda _{2}}\right) ^{2}}%
\times   \label{SIDex} \\
&&\left[ \lambda _{2}\left( e^{\lambda _{2}}-1\right) \left( e^{\lambda
_{1}}+1\right) -\right.   \nonumber \\
&&\left. \lambda _{1}\left( e^{\lambda _{1}}-1\right) \left( e^{\lambda
_{2}}+1\right) \right]   \nonumber
\end{eqnarray}
with 
\begin{equation}
\lambda _{1,2}=-\frac{1}{2}\frac{\tau _{D}}{\tau _{T}}\left( 1\pm \sqrt{1+4%
\frac{\tau _{T}^{2}}{\tau _{D}\tau _{d}}}\right)   \label{lambda}
\end{equation}
Here, $\tau _{D}=L^{2}/D$ is the diffusion transit time, i. e. the time
required for a particle to travel a distance $L$ due to diffusion, with $D$
being the diffusion coefficient, related to the mobility through Einstein's
relation, $D/\mu =k_{B}T/q$. By considering the right hand side of Eq.(\ref
{SID}), we note that the first term corresponds to the expected Nyquist
noise contribution, while the second term is due to finite size effects
controlled essentially by the interplay between the values of the dynamic
transit time and the dielectric relaxation time\cite{gomila00}. The
continuous line in Fig. \ref{Fig3} corresponds to the theoretical results
obtained from the model after using the parameters of the device under test.
The agreement between theory and experiments is within experimental
uncertainty, what is remarkable {\em in view of the absence of any
adjustable parameters}. In particular, the anomalous cross-over between
Nyquist and shot noise found in the experiments is well reproduced by the
theory. Therefore, we conclude that the shot noise observed in our sample is
due to the inelastic drift and diffusion of the carriers under the condition
that the long range Coulomb interaction is not affecting the current
fluctuations, and is not related to a potential energy barrier or to the
absence of inelastic scattering. According to the present view, the fact
that shot noise is not observed in macroscopic resistors made of good
conductors (e.g. metals) is due to the strong correlations induced by the
long range Coulomb interaction in these samples that inhibits the pile up of
charge carriers under realistic experimental conditions, rather than due to
the presence of inelastic scattering processes, as sometimes claimed in the
literature.\cite{blanter00,shimizu92,dejong95} It is worth noting that the
absence of long range Coulomb correlations is also at the basis of the
presence of shot noise in vacuum tubes and ballistic diodes under saturation
\cite{gomila02a} and Schottky barrier diodes\cite{Gomila00b} or number
fluctuations in non-degenerate Fermi gases\cite{Wilson01}.

The strong dependence on temperature of the resistivity of semi-insulating
CdTe\cite{strauss97} allows us to perform experiments for different sample
resistivities by simply varying the sample temperature. Figure \ref{Fig4}
reports the forward measured $I-V$ characteristics for the device under
study at $T=280$ K and $T=300$ K (for an easier comparison the results at 
$T=323$ K are also displayed). At the different temperatures the device still
exhibits a linear behaviour in a broad range of applied bias (only at $T=280$
K minor deviations from linearity are observed at the highest voltages
probably due to some injection effects from the contacts).The lines in Fig. 
\ref{Fig4} represent a best linear fit to the experimental results, giving
for the resistances $R_{300K}=1.6$ G$\Omega $ and $R_{280K}=10$ G$\Omega $.
The measurements of the current spectral density as a function of frequency
at $T=280$ K and $T=300$ K are found to exhibit the general features of the
spectra reported in Fig. \ref{Fig2}. The values of the spectra at the
intermediate white plateaux are plotted in Fig. \ref{Fig5} as a function of
the current (the results obtained at $T=323$ K are also displayed for an
easier comparison). As seen in the figure, the results for the different
resistivities follow the general trends already described for $T=323$ K (see
Fig. \ref{Fig3}), and, in particular, they confirm the presence of shot
noise in macroscopic resistors. The simple unipolar drift-diffusion model
(lines in Fig. \ref{Fig5}) reproduces the main trends of the experimental
results for the different temperatures in spite of the absence of any
adjustable parameter and of its extreme simplicity. The disagreements
observed in the cross-over region between Nyquist and shot noise are
probably due to generation-recombination noise, here neglected.\cite
{Gomila02}

In summary, we have presented experimental evidence of the existence of shot
noise in linear macroscopic resistors, and proved that shot noise can be
observed in electronic devices even in the absence of an internal potential
energy barrier or in the presence of inelastic scattering processes. Present
findings provide support for the existance of an additional physical
situation in which shot noise can be present in electronic devices.

Acknowledgments. Prof. A. Cavallini of Bologna University is thanked for
valuable information provided on the electrical properties of
semi-insulating CdTe. Partial support from the Spanish MCyT through the
Ram\'{o}n y Cajal Program, the European Comission through project No.
IST2001-38899 and the Italy-Spain joint action No HI02-40 is gratefully
acknowledged.

\begin{figure}[tbp]
\caption{Current-voltage characteristics of the CdTe resistor at $T=323$ K
(filled circles). A best linear fit to the experiments (dashed line) gives a
resistance $R_{323K}=0.233 G\Omega$.}
\label{Fig1}
\end{figure}

\begin{figure}[tbp]
\caption{Spectral density of the current fluctuations as a function of
frequency for different values of the electric current at $T=323$ K.}
\label{Fig2}
\end{figure}

\begin{figure}[tbp]
\caption{Experimental values of the current spectral density at the white
plateaux as a function of the current at $T=323$ K (filled circles). The
continous line corresponds to the theoretical prediction of the unipolar
drift-diffusion model. The dotted line corresponds to the standard coth-like
behaviour}
\label{Fig3}
\end{figure}

\begin{figure}[tbp]
\caption{Current-voltage characteristics for the CdTe resistor at three
different temperatures: $T=280$ K (filled triangles), $T=300$ K (filled
circles) and $T=323$ K (filled squares). The lines correspond to a best
linear fit to the experiments: continuous line $T=323K$, dashed line $T=300$%
, and dotted line $T=280$.}
\label{Fig4}
\end{figure}

\begin{figure}[tbp]
\caption{Experimental values of the current spectral density at the white
plateaux as a function of current at three different temperatures: $T=280$ K
(filled triangles), $T=300$ K (filled circles) and $T=323$ K (filled
squares). The lines correspond to the theoretical predictions of the
unipolar drift diffusion model: continuous line $T=323K$, dashed line $T=300$%
, and dotted line $T=280$.}
\label{Fig5}
\end{figure}

                                                                                                                                                                                                                                                                                                            

\begin{references}
\bibitem{sampietro00}  M. Sampietro, G. Ferrari and M. Bertuccio, J. Appl.
Phys., {\bf 87}, 7583 (2000). We note that the value reported in this
reference for the mobility (25 cm$^{2}$/(Vs)) was calculated assuming that
the cut-off frequency is equal to the inverse of the transit time. According
to the calculations in Ref. \cite{gomila00} the cut-off frequency is half
the inverse of the transit time, where the value of 50 cm$^{2}$/(Vs) was
used for the mobility in the present Letter.
\end{references}
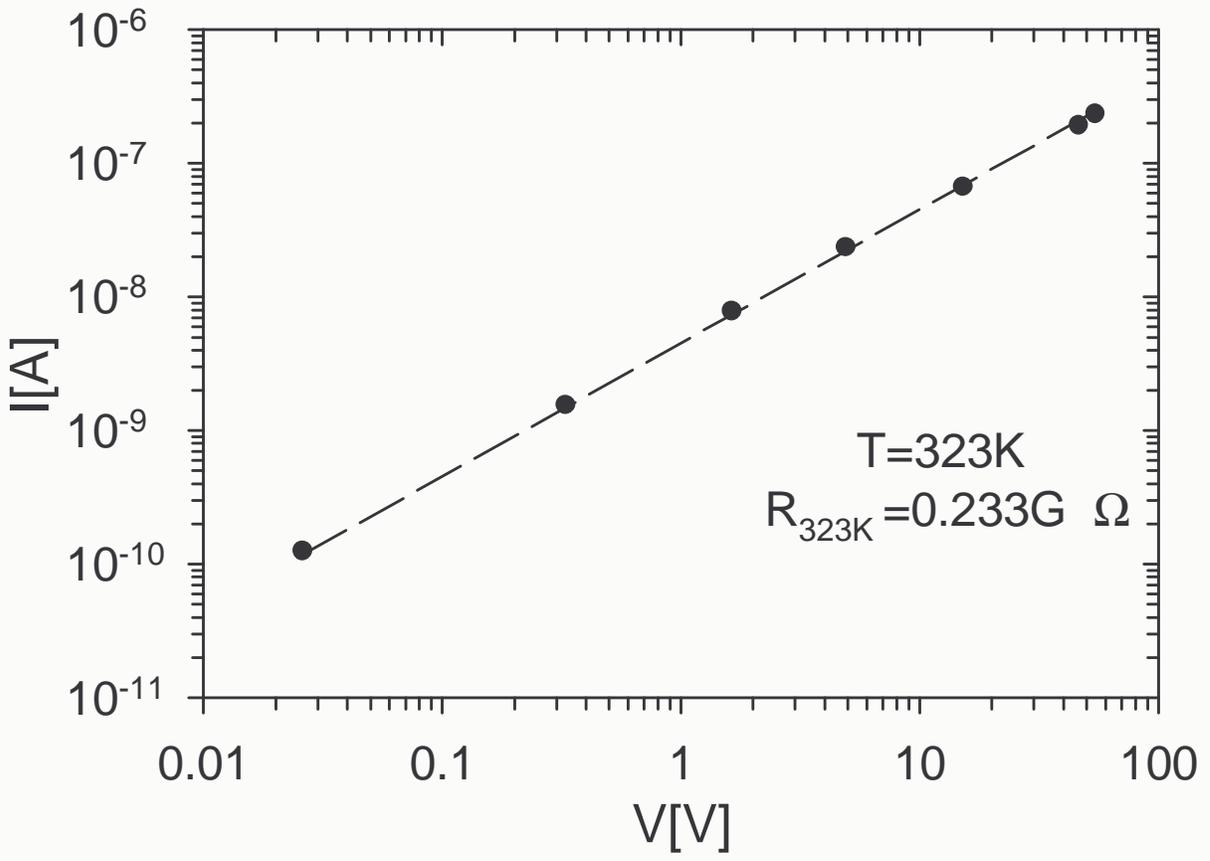
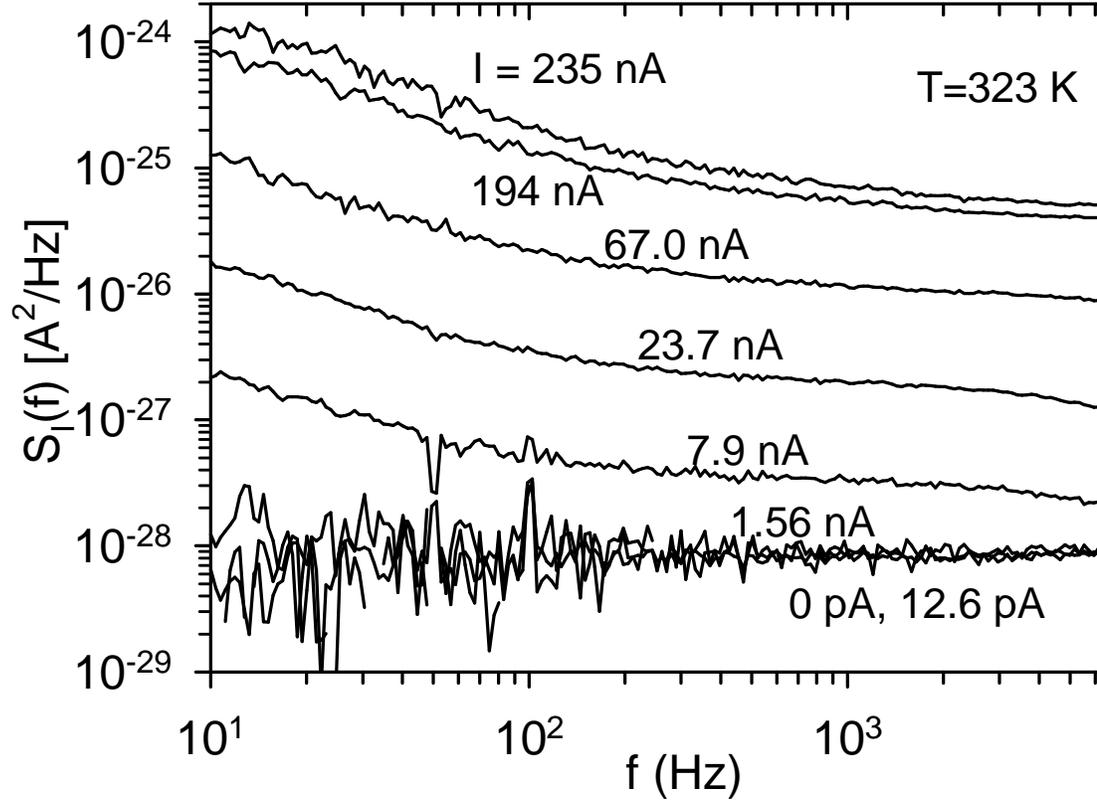
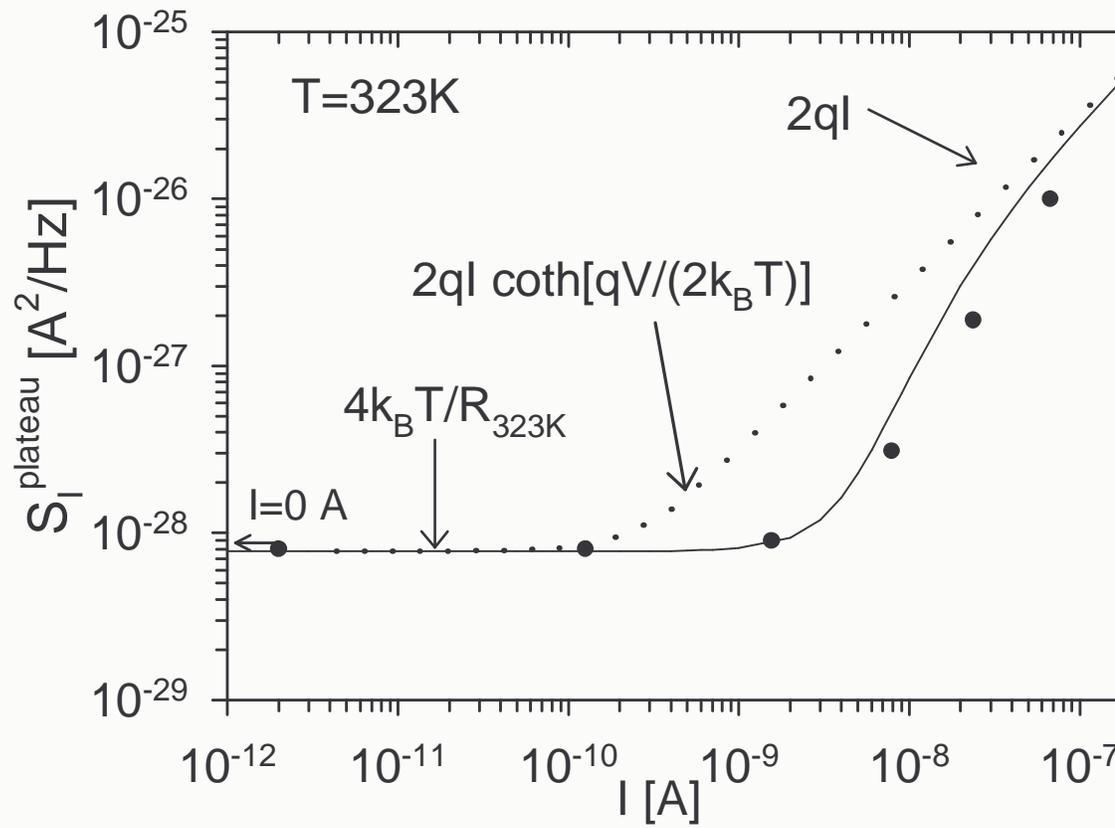
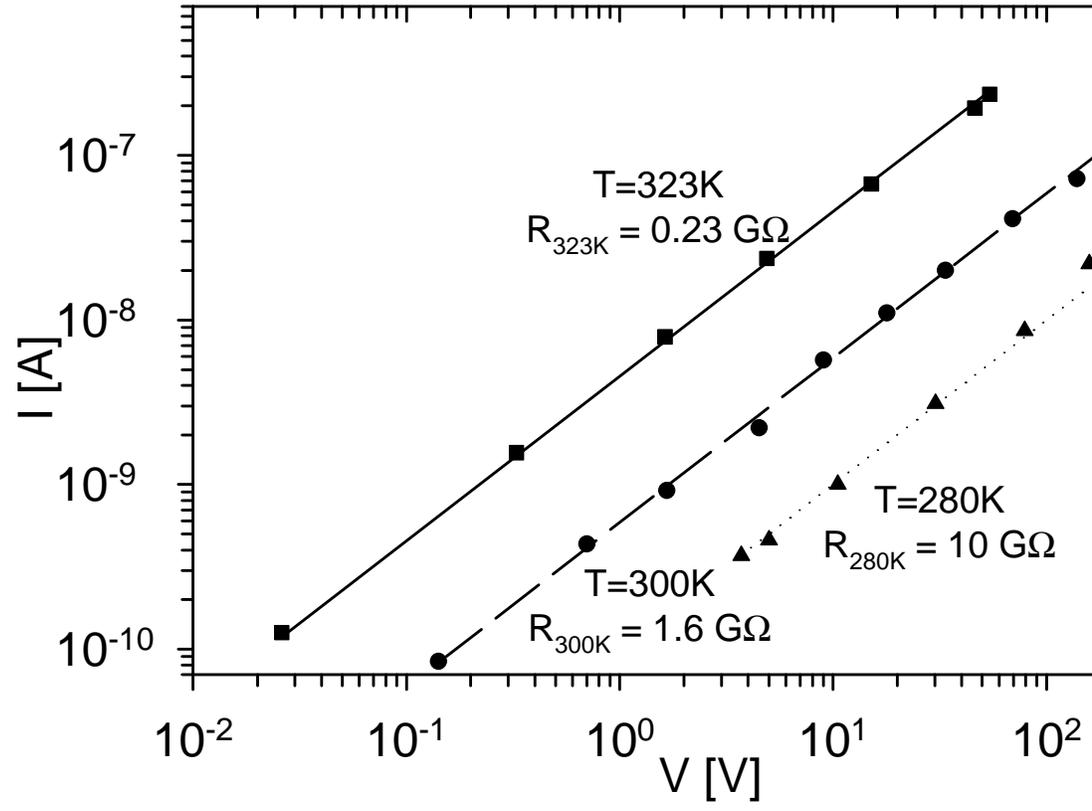
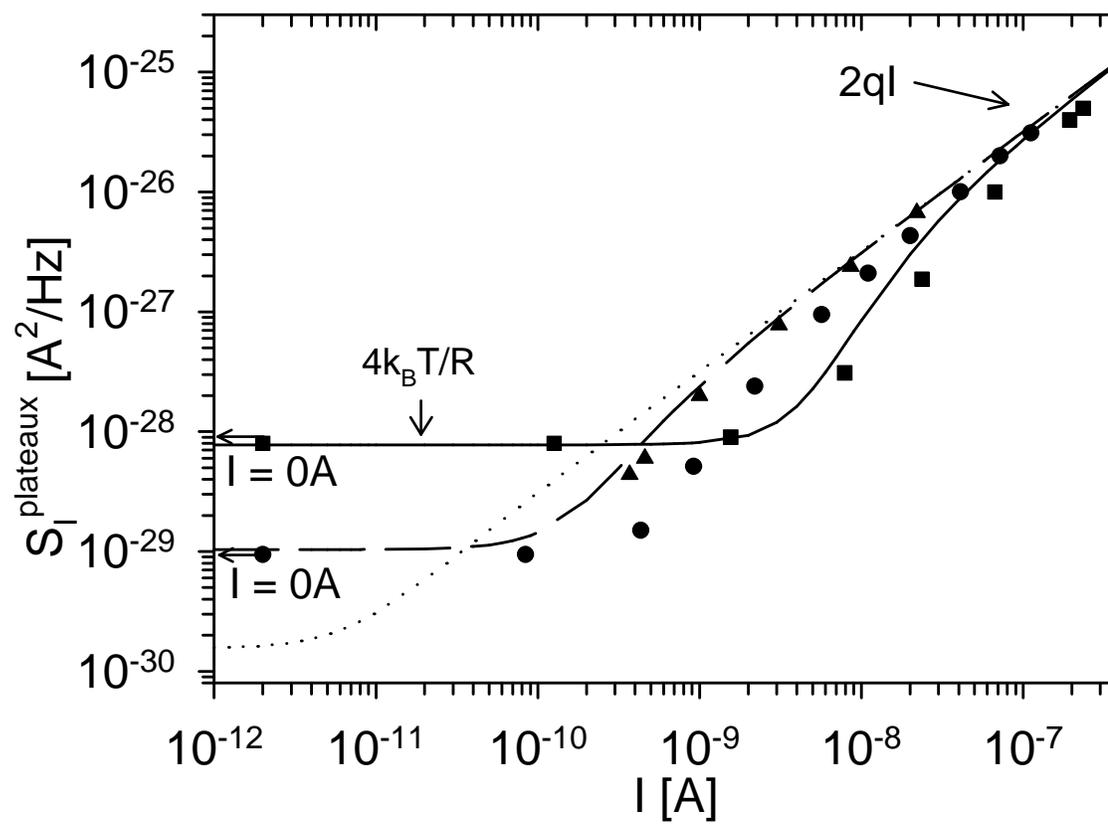
\end{document}